\input{aipcheck}

\documentclass[
    ,final            
  ]
  {aipproc}

\layoutstyle{6x9}

\begin{document}

\title{Mass loss and supernova progenitors}

\classification{97.10.Cv, 97.10.Me, 97.60.Bw}
\keywords      {stars: evolution, supernovae: general, supernovae: 2002ap, 2006jc.}

\author{John Eldridge}{
  address={Astrophysics Research Centre, Physics Building, Queen's University, Belfast, BT7 1NN, UK.}
}

\begin{abstract}
We first discuss the mass range of type IIP SN progenitors and how the upper and lower limits impose interesting constraints on stellar evolution. Then we discuss the possible implications of two SNe, 2002ap and 2006jc, for Wolf-Rayet star mass-loss rates and long Gamma-ray bursts.
\end{abstract}

\maketitle


\section{Introduction}

Core-collapse supernovae (SNe) are the violent deaths of stars more massive than $\approx~7.5 {\rm M}_{\odot}$. They occur when nuclear burning reactions or electron degeneracy-pressure can no longer support the core against gravitational collapse. Either a neutron star or black hole is formed from the collapsing iron core and the outer layers of the star explode in a violent display. The nature of this display depends strongly on the final state of the progenitor star and the circumstellar medium; because there are many paths of stellar evolution the are many types of SNe.

SNe are classified according to their observed spectra and lightcurves. The first differentiation is made by the absence or presence of hydrogen in a spectrum. If hydrogen emission lines exist then a SN is type II and type I otherwise. For type II SNe there are four subgroups: IIP when there is a plateau to the lightcurve lasting a few months. These are the most common type of SN. Type IIL have a linear decay to their lightcurve, IIn have narrow hydrogen emission lines in their spectrum and IIb initially appear to be type II until after a short time the hydrogen lines disappear and the SN becomes type Ib.

The hydrogen free type I SNe have three separate subtypes. Type Ia are thought to be thermonuclear explosions of Chandrasekhar mass white dwarfs and are not considered further here, type Ib have helium lines in emission while type Ic have no helium lines. 

Single star models predicts that the type II SNe will be the result of stars between about $7$ to $27M_{\odot}$ \citep{H03,ETsne,arend} as these retain their hydrogen envelopes. Stars more massive than this lose all their hydrogen via stellar winds and therefore give rise to type Ib/c SNe. The only way to confirm this mapping is to observe the progenitors of SN. This is achieved by searching telescope archives to discover pre-explosion images.

While the progenitors of SNe 1987A and 1993J where detected both were in nearby galaxies and both rare and unusual SNe. It was not until 1999 when the HST archive covered enough galaxies at sufficient depth and resolution that it was possible to start looking for progenitors of SN in a large number of galaxies at distances up to 20 megaparsecs. The first detection for the most common, type IIP, SN was for 2003gd \citep{2004Sci...303..499S}. This confirmed that their progenitors were red supergiants. With eight years of observations there are currently 32 SNe with useful pre-explosion images, 18 of these are type IIP (6 detections) and the remainder being types IIb and Ib/c. There are no detections of type Ib/c progenitors and with the growing number of the non-detections there is growing evidence that our standard view of mass loss during the late stages of evolution may be incorrect.

In this proceedings we briefly highlight some of the main conclusions the mass range inferred for type IIP progenitors. We then discuss two interesting cases of two type Ib/c SNe, 2002ap and 2006jc that provide very stringent limits on the evolution of the most massive stars.

\section{The population of type IIPs.}

With the sample of 18 type IIP detections and non-detections it is possible to begin to characterise the population of the progenitors \citet{compilation}. The main important result for their study is that by using a maximum likelihood analysis it is possible to infer the minimum and maximum initial masses for type IIP SN progenitors. The initial mass range is between $7.5$ to $16.5M_{\odot}$, however the error bars are large and the range could be as large as $6$ to $18{\rm M}_{\odot}$. The initial mass depends strongly on the mixing and mass-loss scheme of the stellar models use to estimate an initial mass from a final luminosity. To remove this systematic it is better to work out the range of final helium core masses which is approximately $1.9$ to $6M_{\odot}$ in the STARS stellar models \citep{ETsne}.

The minimum mass can be used to constrain models of convection in stellar models. Most models with helium cores at the lower end of this range experience second dredge-up and become AGB stars. In fact the stellar models used in this work do experience second dredge-up and we use the pre-dredge up models. Therefore something is required to prevent these stars from becoming AGB stars. It is possible for the most massive AGB stars to undergo SN, however their observational signature in pre-explosion images would be quite different to the red supergiants observed to date as they are cooler and therefore more luminous at infra-red wavelengths \citep{EMS07}.

The maximum mass is due to one of two factors, because stars above this limit have lost most (or all) of their hydrogen and produce another type of SN or it is because the cores are massive enough to form a black hole and this also leads to another type of SN. In reality it is probably a combination of these factors. However the black hole explanation could be favoured as the inferred helium core mass is similar to that which is expected to produce a black hole rather than a neutron star at core-collapse \citep{H03,ETsne}.

\section{But what about other types?}

With only one detection for the other SN types there is little that can be said. If the upper limits that have been obtained on the progenitors' luminosity are compared to the luminosity of observed Wolf-Rayet stars, the suspected progenitors, it is apparent that the pre-explosion in all but one cases are not deep enough to have revealed the progenitors. For a detection it is a waiting game to determine for type type Ib/c SN to occur nearby and have deep pre-explosion images.

The culprits for type Ib/c progenitors are Wolf-Rayet (WR) stars. These are evolved massive stars that have completed core hydrogen burning and lost (or in the process of losing) their hydrogen envelopes and are naked helium stars. These stars are also the preferred progenitors of long Gamma-ray bursts \citep{grbreview}.

Despite this there are two interesting pre-explosion images for type Ib/c SNe. SN 2002ap, because the limit is so low that we were only just unable to detect the progenitor and are able to rule out single stars for the progenitor. While before SN 2006jc a luminous outburst transient was detected two years prior to the SN and this may have interesting consequences for our understanding of stellar-wind mass-loss.

\subsection{2002ap}

\citet{crockett} used previously unused deep pre-explosion images to reexamine the progenitor of SN 2002ap. The limit is the most stringent to date on any type Ib/c progenitor. They were able to rule out all standard single-star models are suggest that the most likely progenitor was a binary. 

\begin{figure}
  \includegraphics[height=.3\textheight]{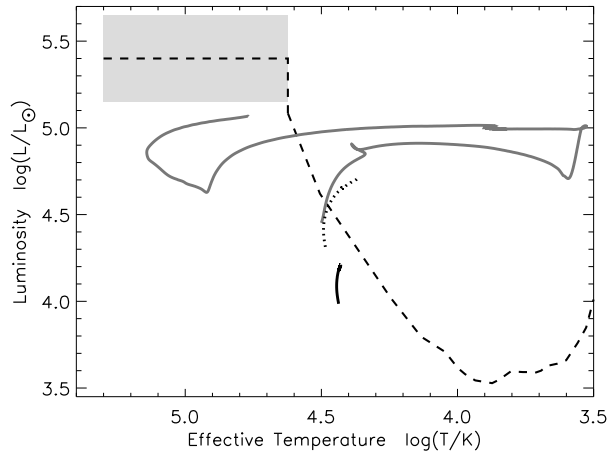}
  \caption{Theoretical Hertzsprung-Russell diagram. The dashed line is the luminosity limit from the non-detection of any object in the pre-explosion image, the grey box represents the uncertainty in the limit for WR stars \citep{crockett}. The solid grey line is the evolution for a $17M_{\odot}$ star which has its hydrogen envelope removed in a binary interaction. The solid black line is the evolution of a $11.9M_{\odot}$ binary companion and the dotted line is the evolution of a $15.3M_{\odot}$ binary companion. At the time of the primary SN the latter companion would have been observed while the former would have remained undetected.}
  \label{02ap1}
\end{figure}

Figure \ref{02ap1} shows the luminosity limit derived from the pre-explosion images on a theoretical Hertzsprung-Russell diagram. The grey stellar track is of a $17{\rm M}_{\odot}$ star that has its hydrogen envelope stripped by interaction via Roche-Lobe Overflow in a binary. The final stellar model has a final mass or $5{\rm M}_{\odot}$ that agrees with the mass inferred from modelling the SN lightcurve by \citet{Mazzali}. Also shown on the figure are the tracks of possible companion stars the mass of any companion star can also be limited by the pre- and post- explosion images. The mass of the companion star must have been less than $\approx 14M_{\odot}$. Any mass transfer must have been quite inefficient otherwise the secondary would have accreted a large amount of mass and be visible in the pre-explosion image.

However there is a problem with this model, there is a large amount of helium in the progenitor model. While the amount of helium required for a type Ib SN is uncertain, over $0.5M_{\odot}$ should produce the signature of helium lines in the SN spectrum. The binary model has around $1M_{\odot}$ of helium in the envelope. Therefore there is some uncertainty in whether this is a reasonable progenitor model.

There are solutions, one is that the star was more massive and underwent more severe stripping during the binary interaction. Another is that we underestimate the strength of stellar winds of Wolf-Rayet stars. Increase the mass-loss rates of a star by a factor of 2 - 3 would reduce the final mass of helium and therefore produce the required type Ic SN. A source of increased mass-loss rates could be rapid rotation [REF].

An alternative is that the mass-loss rates of WR stars are underestimated. This would be possible if the mass-loss rates are enhanced for only a short time of evolution. A possible mechanism has been suggested by \citet{petrovic} who discuss how helium star envelopes can become inflated. Inflated helium envelopes are caused by a bump in the opacity which causes the envelopes to become extended and the density profile of the envelope to invert so density increases with radius from the core. \citet{petrovic} suggest that this inflation may not be physical and note that by increasing the mass-loss rates it is possible to remove the inflated envelope and density inversion.

\begin{figure}
  \includegraphics[height=.3\textheight]{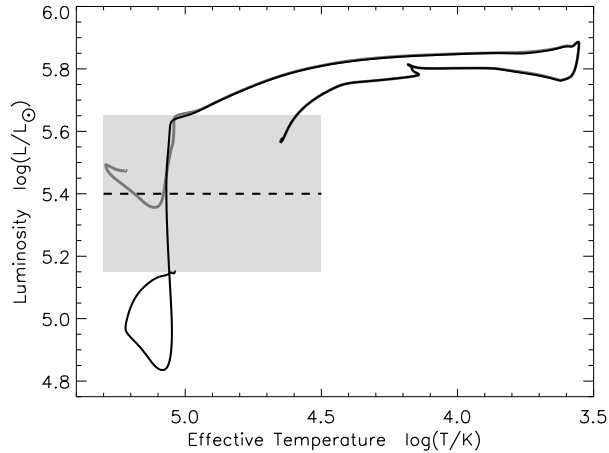}
  \caption{Theoretical Hertzsprung-Russell diagram of the evolution of initially $50M_{\odot}$ stars. The dashed line is the luminosity limit from the non-detection of a WR star the pre-explosion image, the grey box represents the uncertainty. The solid grey line is for a standard WR model \citep{EV06} and the solid black line a model using the mass-loss rates of \citep{petrovic}. }
  \label{02ap2}
\end{figure}

Figure \ref{02ap2} shows a standard stellar model with the mass-loss rates of \citet{NL00} and a model which replaces these with the mass loss rates of \citet{petrovic} when the helium envelope becomes inverted. The figure shows that during the nitrogen rich WN evolution much more mass is lost than in the standard case. Further more most of the lost material is helium. The final model retains only $0.25M_{\odot}$ and the mass again agrees with the analysis of \citet{Mazzali}.

With one non-detection it is not possible to decide if all WR mass-loss rates need to be increased but if the number of non-detections increase then the situation may begin to become more serious and the mass-loss rates will have to be closely examined. The only other scenario is that WR stars produce large amounts of dust a few years before core-collapse. This would reduce their apparent luminosity in pre-explosion images.

\subsection{2006jc}

\citet{pastorello} and \citet{folye} describe a most unusual SN. SN 2006jc was discovered on 9th October 2006 in the galaxy, UGC4904. \citet{pastorello} found that it was spatially coincident with a bright optical transient that occurred in 2004. The SN itself was classified as a type Ib-n due to narrow helium lines in the spectra. The current interpretation is that the narrow helium lines are due to helium-rich material ejected by the star in a dramatic mass-loss episode that was observed as the optical transient. Then two years after this episode the star exploded as a type Ic SN and therefore the progenitor star had be stripped of helium, otherwise broad helium emission lines would have been observed in combination with the narrow lines.

The problem with this interpretation is that if the progenitor was a single star then our understanding of Wolf-Rayet stars and their winds must be revised. The only single stars known to produce similar bright optical transients are LBV stars \citep{lbv1,lbv2}. However these stars tend to retain some hydrogen which would have been observed in the SN spectrum. It is conceivable that there are transition objects between LBVs and WR stars that would lead to the observed evolution for the progenitor of SN2006jc. These objects may be rare to get the right amount of mass-loss to remove all helium just before core-collapse. Although such objects would be more common at lower metallicities.

An alternative to the LBV-like WR star is that the transient and mass-loss was due to a pair-production instability causing a dramatic mass-loss episode a few years before a normal core-collapse SN \citep{langerinprep}. The models of \citet{heger02} show that massive stars with little mass loss will experience such outbursts prior to core-collapse. The upper metallicity limit is uncertain. It is possible to estimate from stellar models to be below SMC metallicity.

\begin{figure}
  \includegraphics[height=.3\textheight]{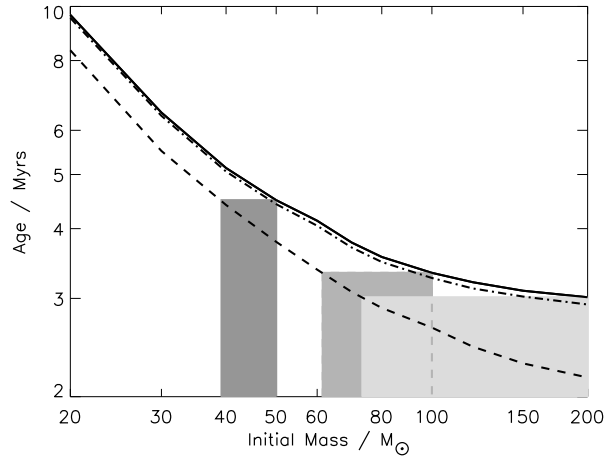}
  \caption{The lifetimes for massive single stars. The solid line is the total lifetime versus initial mass. The dashed line is the time of the end of core hydrogen burning and the dotted-dashed line is the time of the end of core helium burning. The three shaded regions show the range of secondary masses which have completed core hydrogen burning and are LBV candidates by the time the primary (the high mass edge of the region) experiences a SN.}
  \label{06jc}
\end{figure}

There is final a possibility with an observed analogue, the may have been progenitor a binary with a WR star and an evolved LBV star. The outburst was produced by the LBV star while the WR star exploded to produce the type Ic SN. There is one similar system in the SMC that has been observe to undergo outburst and at some point in the future may lead to a similar SN \citep{smclbvwrbin}. These systems are not as rare as might be first thought. If we assume LBV evolution happens after core-hydrogen burning is complete then any star which has a secondary companion that has completed core-hydrogen burning could be a possible LBV-WR system. Figure \ref{06jc} shows how the hydrogen burning lifetime compares to the total lifetime for stars with different initial masses. It is possible to see that the more massive stars can have a wider range of secondary masses for this kind of evolution. If we assume the mass ratio of binaries ($q=M_{\rm secondary}/M_{\rm primary}$) has a flat distribution and is independent of primary mass then 60\% of binaries with a $200M_{\odot}$ primary might be LBV-WR systems, while this reduces to only 20\% for a primary initially $50M_{\odot}$.

The only way to distinguish between these three plausible models will be the rate of these events and the metallicities of the host galaxies. The binary scenario will be only weakly metallicity dependent, the pair-production outburst will be concentrated to low metallicities while the single star LBV/WR has an unknown metallicity dependence but if it is related to inflation of the WR star then is will be concentrated at higher metallicities. 

The total rate of type Ib-n SN is $< 4$\% of all type Ib/c SN. The rate of GRBs as a fraction of all Ib/c SNe is between 0.1 to 1\% [REF]. It is possible therefore that some GRBs may occur with 2006jc-like SNe. If this is the case then the circumburst medium inferred from the optical afterglow would be a constant density medium due to the changes in mass-loss rate and wind speed. There are a number of GRBs where this has been observed and is another possible solution when the CBM is a constant density rather than that expected of a free-wind density profile \citep{vanmarle,eldridge}. 

\section{Conclusions}

While the progenitors of type IIP SNe are becoming well understood there is still great uncertainty over the progenitors of other SN types. It is becoming apparent that our understanding of WR mass-loss maybe incorrect and one possible reason that we do not observe more WR stars is they lose more mass than currently thought and that some of this mass-loss may be in luminous outburst such as the one that proceeded SN 2006jc. Or they produce copious amount of dust in the last few years before core-collapse.

\begin{theacknowledgments}
JJE would like to thank Stephen Smartt, Andrea Pastorello, Seppo Matilla, Mark Crockett and Dave Young for many discussions and making his time at Queen's University Belfast so enjoyable.
\end{theacknowledgments}

\bibliographystyle{mn2e}

\bibliography{eldridge2}

\end{document}